\documentclass[preprints,article,accept,pdftex,moreauthors]{mdpi} 

\firstpage{1} 
\pubvolume{1}
\issuenum{1}
\articlenumber{0}
\pubyear{2023}
\copyrightyear{2023}
\datereceived{ } 
\dateaccepted{ } 
\datepublished{ } 
\hreflink{https://doi.org/} 
\pdfoutput=1 

\Title{Causal and Stable Relativistic Hydrodynamic Fluctuations}

\TitleCitation{A Theory of Relativistic Fluctuations}

\Author{Nicki Mullins $^{1}$*\orcidA{}, Mauricio Hippert $^{1}$, Jorge Noronha $^{1}$ and Lorenzo Gavassino$^{2}$}

\AuthorNames{Nicki Mullins, Mauricio Hippert, Jorge Noronha and Lorenzo Gavassino}

\AuthorCitation{Mullins, N.; Hippert, M.; Noronha, J.; Gavassino, L.}

\address{%
$^{1}$ \quad Illinois Center for Advanced Studies of the Universe, Department of Physics, 
University of Illinois at Urbana-Champaign, Urbana, IL 61801, USA \\
$^{2}$ \quad Department of Mathematics, Vanderbilt University, Nashville, TN, USA}

\corres{Correspondence: nickim2@illinois.edu}

\abstract{When two nuclei collide close to the speed of light, a fluid state known as the quark-gluon plasma is formed. Attempts to understand the dynamics of this fluid have generated significant research into dissipative relativistic fluid dynamics. The fluctuation-dissipation theorem implies that any dissipative dynamical system will also experience thermal fluctuations; however, such fluctuations are not typically included in the modeling of the quark-gluon plasma. This work discusses a new method of determining whether a hydrodynamic framework is consistent with thermal fluctuations. We develop a new method for calculating the noise correlator of relativistic hydrodynamic systems and apply it to Israel-Stewart theory in a general hydrodynamic frame.}


\begin{document}


\section{Introduction}

Relativistic hydrodynamics has had great success modeling the matter formed in heavy-ion collisions. However, the fluctuation-dissipation theorem implies that any dissipative system should also experience thermal fluctuations \cite{Callen:1951vq, Kubo:1957mj}. These fluctuations are not typically included in the modeling of relativistic heavy-ion collisions, though they should be important both near the expected QCD critical point and in small systems. 

In this work, we discuss how causality and stability constraints can be enforced at the level of thermal fluctuations through the newly developed information current \cite{Gavassino:2021kjm}. We then use the information current to formulate a theory of relativistic fluctuations \cite{Mullins:2023tjg, Mullins:2023ott, Gavassino:2024vyu}, similar to that of Fox and Uhlenbeck for nonrelativistic physics \cite{doi:10.1063/1.1693183}. Using this new formulation, we determine the noise correlators for conformal Israel-Stewart theory in a general hydrodynamics frame \cite{Noronha:2021syv} and see that new structures appear in the energy-momentum correlator. 

\section{Materials and Methods}
\label{Sec:Methods}
 
To model the inclusion of thermal fluctuations in relativistic hydrodynamics, we include a stochastic source in the linearised equation of motion: 
\begin{equation}
    \hat{D}_{AB} \delta \phi^B = \xi_A ,
\end{equation}
for some linear differential operator $\hat{D}_{AB}$. Here, capital letter indices run over the space of dynamical variables, $\delta \phi^A$ is a vector containing all the dynamical variables of the system (linearised about equilibrium), and $\xi_A$ is the stochastic source. The distribution from which this stochastic source is drawn is determined such that the correct equilibrium probability distribution is realized.

In nonrelativistic physics, the equilibrium probability distribution at a given time is determined from the free energy. However, when the free energy is constructed in a Minkowski spacetime, the notion of ``time'' is relative. Thus we need to foliate spacetime into a set of spacelike Cauchy hypersurfaces $\Sigma_{\tau}$ with past-directed normal vectors $n^{\mu}$. The free energy within a given hypersurface is then $\delta \Omega = T \int_{\Sigma} d\Sigma n_{\mu} \left( -\delta s^{\mu} - \alpha^*_I \delta J^{\mu I} \right)$, where ``$\delta$'' denotes the variation from equilibrium, $s^{\mu}$ is the entropy current, $\alpha_I^*$ are the equilibrium values of the thermodynamic conjugates, and $J^{\mu I}$ are the conserved currents. This motivates the definition of the information current \cite{Gavassino:2021kjm},
\begin{equation} \label{Eq:information_current}
    E^{\mu} = -\delta s^{\mu} - \alpha_I^* \delta J^{\mu I} .
\end{equation}
In terms of the information current, the equilibrium probability distribution is given by,
\begin{equation} \label{Eq:equilibrium_distribution}
    w_{\mathrm{eq}}[\delta \phi] \sim e^{-\int_{\Sigma} d\Sigma n_{\mu} E^{\mu}} .
\end{equation}

Writing this distribution in terms of the information current is particularly useful because it has been shown that the information current is connected to stability and causality in relativistic theories \cite{Gavassino:2021kjm}. In particular, any relativistic system for which 
\begin{enumerate}
    \item $n_{\mu} E^{\mu} \geq 0$ for any timelike, past-directed $n^{\mu}$,
    \item $n_{\mu} E^{\mu} = 0$ if and only if $\delta \phi^A = 0$,
    \item $\partial_{\mu} E^{\mu} \leq 0$,
\end{enumerate}
will be linearly causal and stable against fluctuations. These conditions can be summarized as follows: $E^{\mu}$ is timelike, future-directed, vanishes in equilibrium, and the second law of thermodynamics holds. 


\section{Results}

In this section, we consider the stochastic dynamics of conformal relativistic hydrodynamics linearized around the equilibrium state. We will take the only conserved quantity to be the energy-momentum tensor 
\begin{equation}
    T^{\mu\nu} = (\epsilon + \mathcal{A}) \left( u^{\mu} u^{\nu} + \frac{1}{3} \Delta^{\mu\nu} \right) + u^{\mu} Q^{\nu} + u^{\nu} Q^{\mu} + \pi^{\mu\nu} ,
\end{equation}
written in the most general possible form. Here, $\epsilon$ is the energy density, $u^{\mu}$ is the fluid velocity, $\mathcal{A}$ is a dissipative correction to the energy density, $Q^{\mu}$ is the heat flux vector, $\pi^{\mu\nu}$ is the shear tensor, and $\Delta^{\mu\nu}$ is the projector orthogonal to $u^{\mu}$. 

\subsection{Information currents for relativistic hydrodynamics}

The most similar theory to Navier-Stokes that can be causal and stable is BDNK theory \cite{Bemfica:2017wps, Kovtun:2019hdm, Bemfica:2019knx, Hoult:2020eho, Bemfica:2020zjp}. We will therefore start with the information current for this theory. Using the definition of equation \eqref{Eq:information_current}, we find that 
\begin{equation} \label{Eq:BDNK_information_current}
\begin{split}
    E^{\mu}_{\mathrm{BDNK}} = & \, \frac{\epsilon u^{\mu}}{T^3} \delta T^2 + \frac{\epsilon u^{\mu}}{3T} \delta u^{\lambda} \delta u_{\lambda} + \frac{c_V}{3T} \delta T \delta u^{\mu} + \frac{1}{3T} \delta \mathcal{A} \delta u^{\mu} + \frac{u^{\mu}}{T} \delta u^{\lambda} \delta Q_{\lambda} + \frac{1}{T} \delta u_{\nu} \delta \pi^{\mu\nu} \\
    & + \frac{u^{\mu}}{T^2} \delta \mathcal{A} \delta T + \frac{1}{T^2} \delta T \delta Q^{\mu} .
\end{split}
\end{equation}
To understand this result, consider the behavior of the heat flux, $Q^{\mu}$. This part of the information current scales as $T^2 E^{\mu}_{\mathrm{BDNK}} \sim Tu^{\mu} \delta u^{\lambda} \delta Q_{\lambda} + \delta T \delta Q^{\mu}$. For conformal BDNK theory at zero chemical potential, $Q^{\mu} = \tau_Q \left[ (\epsilon + P) u^{\lambda} \partial_{\lambda} u^{\mu} + \Delta^{\mu\lambda} \partial_{\lambda} P \right]$, where $P=\epsilon/3$. The equilibrium probability distribution, equation \eqref{Eq:equilibrium_distribution}, then favors states with large, negative derivatives in pressure and fluid velocity over the equilibrium state. This signals a violation of the conditions for stability against fluctuations presented in section \ref{Sec:Methods}. One can see such a lack of stability through the absence of quadratic terms in the dissipative quantities of equation \eqref{Eq:BDNK_information_current}. If a stochastic hydrodynamic theory is constructed for BDNK theory naively, the resulting noise correlators will not be positive definite, as discussed in \cite{Mullins:2023ott}. However, it has recently been shown that by introducing UV regulators in the information current \cite{Gavassino:2024ufs} it is possible to consistently study fluctuating BDNK theory at the linear level \cite{Gavassino:2024vyu}.

In this work, we consider Israel-Stewart theory \cite{Israel:1979wp, Noronha:2021syv}. Taking a general hydrodynamic frame, the information current is
\begin{equation} \label{Eq:gIS_information_current}
\begin{split}
    E^{\mu}_{\mathrm{gIS}} = & \, \frac{\epsilon u^{\mu}}{T^3} \delta T^2 + \frac{\epsilon u^{\mu}}{3T} \delta u^{\lambda} \delta u_{\lambda} + \frac{c_V}{3T} \delta T \delta u^{\mu} + \frac{1}{3T} \delta \mathcal{A} \delta u^{\mu} + \frac{u^{\mu}}{T} \delta u^{\lambda} \delta Q_{\lambda} + \frac{1}{T} \delta u_{\nu} \delta \pi^{\mu\nu} \\
    & + \frac{u^{\mu}}{T^2} \delta \mathcal{A} \delta T + \frac{1}{T^2} \delta T \delta Q^{\mu} + \frac{u^{\mu}}{2T} \left[ \frac{\tau_{\mathcal{A}}}{4\epsilon \tau_{\phi}} \delta \mathcal{A}^2 + \frac{\tau_Q}{4\epsilon \tau_{\psi}} \delta Q^{\lambda} \delta Q_{\lambda} + \frac{\tau_{\pi}}{\eta} \delta \pi^{\alpha\beta} \delta \pi_{\alpha\beta} \right] ,
\end{split}
\end{equation}
where $\tau_{\mathcal{A}}, \tau_Q, \tau_{\phi}, \tau_{\psi}, \tau_{\pi}$ are new coefficients that appear in relaxation-type equations for the dissipative quantities. For suitable choice of these coefficients, this information current satisfies the conditions introduced in section \ref{Sec:Methods}.

\subsection{Fox-Uhlenbeck approach}
\label{Sec:FU_approach}

The equation of motion of a linear relativistic system can be written in the form
\begin{equation}
    \left( M_{AB}^{\mu} \partial_{\mu} + V_{AB} \right) \delta \phi^B = \xi_A .
\end{equation}
The vector $\xi_A$ will be taken to be normally distributed with zero mean. We then separate out a timelike derivative by decomposing,
\begin{equation}
    \left( M_{AB}^{\mu} n_{\mu} \frac{d}{d\tau} + M_{AC}^{\mu} n_{\mu} F_{CB} \right) \delta \phi^B = \xi_A ,
\end{equation}
where $d/d\tau = -n^{\mu} \partial_{\mu}$, and $F_{AB}$ is a differential operator that contains the derivatives within the hypersurface and $V_{AB}$. We can then express the equation of motion in the form
\begin{equation}
    \left( \delta_{AB} \frac{d}{d\tau} + F_{AB} \right) \delta \phi^B = \tilde{\xi}_A ,
\end{equation}
which is a matrix Langevin equation.

Such linear systems will have a quadratic information current,
$E^{\mu} = \delta \phi^A E_{AB}^\mu \delta \phi^B / 2$. Using the equilibrium probability distribution of equation \eqref{Eq:equilibrium_distribution}, we find that the noise correlator of $\tilde{\xi}$ should have the form 
\begin{equation}
    \langle \tilde{\xi}_A(x) \tilde{\xi}_B(y) \rangle = \left[ F_{AC} (n_{\mu} E^{\mu})^{-1}_{CB} + (n_{\mu} E^{\mu})^{-1}_{AC} F_{CB}^{\dagger} \right] \delta^{(4)}(x-y) .
\end{equation}
Here, we have assumed that the noise correlators are independent of the momentum and frequency in Fourier space. This assumption is valid in all Israel-Stewart like theories. 

\subsection{Fluctuating Israel-Stewart in a general hydrodynamic frame}

Using the approach developed in section \ref{Sec:FU_approach}, we can calculate the noise correlators for Israel-Stewart in a general hydrodynamic frame, with information current given by equation \eqref{Eq:gIS_information_current}. We find that the conservation law $\partial_{\mu} T^{\mu\nu}=0$ receives no stochastic corrections (the corresponding source has zero correlator), while the relaxation equations for the dissipative quantities fluctuate as
\begin{equation}
\begin{split}
    & \langle \xi_{\mathcal{A}}(x) \xi_{\mathcal{A}}(y) \rangle = 4 \epsilon T \tau_{\phi} \delta^{(4)}(x-y) , \\
    & \langle \xi_{Q}^{\mu}(x) \xi_Q^{\nu}(y) \rangle = 4 \epsilon T \tau_{\psi} \Delta^{\mu\nu} \delta^{(4)}(x-y) , \\
    & \langle \xi_{\pi}^{\mu\nu}(x) \xi_{\pi}^{\alpha\beta}(y) \rangle = 2 \eta T \Delta^{\mu\nu\alpha\beta} \delta^{(4)}(x-y) .
\end{split}
\end{equation}
These noise correlators can be used to determine the correlator of the energy-momentum tensor. This result is presented in \cite{Mullins:2023tjg}. Interestingly, by including all possible terms in the energy-momentum tensor rather than taking the Landau frame we find that new structures appear in the energy-momentum correlator. In the Landau frame, this correlator scales with $\Delta^{\mu\nu\alpha\beta}$; however, all possible structures appear in a general hydrodynamic frame. 


\section{Conclusions}

Using the information current, we developed a theory of stochastic relativistic hydrodynamics that is causal and stable against thermal fluctuations. This new approach is applied to Israel-Stewart theory in a general hydrodynamic frame, and it is found that new structures appear in the correlator of the energy-momentum tensor. 

The results in this paper have since been extended to study the properties of the effective action corresponding to a stochastic hydrodynamic theory in \cite{Mullins:2023ott}. In particular, it was found that the fluctuation-dissipation relation can also be determined as a realization of detailed balance through a symmetry of the effective action \cite{Guo:2022ixk, Huang:2023eyz}.



\vspace{6pt} 





\funding{NM and JN are partially supported by the U.S. Department of Energy, Office of Science, Office for Nuclear Physics
under Award No. DE-SC0021301 and DE-SC0023861. MH and JN were partially funded by the National Science Foundation within the framework of the MUSES Collaboration under grant number OAC-2103680. LG is partially supported by a Vanderbilt's Seeding Success Grant.}

\conflictsofinterest{The authors declare no conflict of interest.} 

\begin{adjustwidth}{-\extralength}{0cm}

\reftitle{References}


\bibliography{references}

\PublishersNote{}
\end{adjustwidth}
\end{document}